\documentclass{article}
\usepackage{a4wide,amssymb,amsmath,cite}
\title{The `BRST-invariant' Condensate of Dimension Two in QCD}
\author{B. M. Gripaios\thanks{b.gripaios1@physics.ox.ac.uk}
\\\emph{Department of Physics - Theoretical Physics, University of Oxford,} 
\\ \emph{1, Keble Road, Oxford, OX1 3NP,  UK}
\\
\\ OUTP-03 05P}
\date{February 3, 2003}
\begin{document}
\bibliographystyle{h-elsevier2}
\maketitle
\begin{abstract}
The status of the `BRST-invariant' condensate of mass dimension two in QCD is explained. The condensate is only invariant under an `on-shell' BRST symmetry which includes a partial gauge fixing. The on-shell BRST symmetry represents the residual gauge symmetry under gauge transformations which preserve the partial gauge fixing. The gauge-invariant operators which correspond to the BRST-invariant condensate are identified in the Lorentz and maximal Abelian gauges and are shown to be invariant under the residual gauge transformations.


\end{abstract}

\section{Introduction}
It is by now well known that vacuum condensates provide signatures of the non-perturbative dynamics of QCD. The quark-antiquark condensate acts as an order parameter for chiral symmetry breaking, the gluon condensate is related to the QCD trace anomaly, and the topological susceptibility gives the $\eta'$ its large mass via the $U(1)$ axial anomaly. Condensates of mass dimension two are also believed to be important. They occur in operator product expansions for gluon propagators \cite{Lavelle:1988eg,Lavelle:1992yh,Gubarev:1999ie,Boucaud:2000ey,Boucaud:2000nd,Boucaud:2001st,Boucaud:2002nc,Boucaud:2002jt,Kondo:2002cy} and may be important for confinement \cite{Schaden:1999ew,Kondo:2000ey,Gubarev:2000nz,Gubarev:2000eu,Verschelde:2001ia,Dudal:2002aj,Dudal:2002xe}. The drawback is that they cannot be discussed in a gauge-invariant fashion, since there are no \emph{local} gauge-invariant quantities of mass dimension two. Calculations are thus dependent on the choice of gauge.

Recently, in the BRST \cite{Becchi:1975md,Becchi:1976nq,Tyutin:1975qk} formulation of QCD, a mixed vacuum condensate of mass dimension two containing both gluons and ghosts has been proposed \cite{Kondo:2001nq,Kondo:2001tm,Kondo:2002xn}. It is claimed that the condensate is BRST-invariant `for any gauge' and moreover that it is of ghost number zero and BRST-inexact. It is thus a non-trivial element of the BRST co-homology at ghost number zero. Now there is a well established one-to-one correspondence between the BRST co-homology at ghost number zero and the set of classical physical observables, that is the gauge-invariant functions \cite{Henneaux:1992ig}. The following question thus arises: if the mixed condensate is indeed BRST-invariant, to what classical observable does it correspond? This presents us with a paradox, for there are no (local) gauge-invariant functions of mass dimension two.

In this letter, the paradox is resolved by considering the subtleties of gauge fixing in the extended phase space of the BRST formalism. Whilst the BRST symmetry in the extended phase space exists independently of the choice of gauge (as it must, since it is equivalent to gauge invariance), the action does not. So the action, and the dynamics, are inevitably gauge-fixed. The construction of the BRST-invariant mixed condensate \cite{Kondo:2001nq} requires one to eliminate the auxiliary (or Takanishi--Lautrup) field via its equation of motion. Thus the dynamics are invoked, and the gauge is partially fixed. The mixed condensate is thus not strictly BRST-invariant. It is invariant under an `on-shell' BRST symmetry corresponding to a partial gauge fixing. What is the classical observable corresponding to the on-shell BRST-invariant condensate? It is a function of mass dimension two which is gauge invariant under the restricted class of gauge transformations which preserve the partial gauge fixing. This is demonstrated explicitly for both classes of gauge considered in the original work, \emph{viz.} the (generalized) Lorentz \cite{Curci:1976bt,Baulieu:1982sb,Baulieu:1985tg} and maximal Abelian \cite{'tHooft:1981ht,Kronfeld:1987vd,Kronfeld:1987ri,Kondo:1998pc,Kondo:1998nw} gauges.

In the next section we recall the details of the BRST transformation in $SU(N)$ gauge theory and discuss the correspondence between elements of the BRST co-homology at ghost number zero and classical observables. 
Then, in section \ref{gfix}, we discuss the form of the action and the inherent gauge fixing. In section \ref{cond} we introduce the mixed condensate and extract the corresponding classical observables. We then show that the classical observables are invariant under the restricted set of gauge transformations preserving the partial gauge fixing. 
\section{The BRST symmetry in QCD}
\label{BRST}
For the pure $SU(N)$ gauge theory, the gauge transformations of the Lie algebra-valued gauge fields $A_{\mu}=A_{\mu}^{A} T^A$ are $A_{\mu} \rightarrow U(A_{\mu} + \partial_{\mu})U^{\dag}$, where $U \epsilon SU(N)$. 
The BRST transformation can be written in the form
\begin{align}
\label{BRSTeq}
\delta A^{A}_{\mu} &= \partial_{\mu} C^{A} + g f^{ABC} A^{B}_{\mu} C^{C}, \nonumber \\
\delta C^{A} &= -\frac{1}{2} g f^{ABC} C^{B}C^{C}, \nonumber \\
\delta \bar{C}^{A} &= ib^{A}, \nonumber \\
\delta b^{A} &= 0,
\end{align}
where $b, C$ and $\bar{C}$ are the auxiliary fields, ghosts and antighosts respectively. The BRST operator $\delta$ has a left action, increases the ghost number by unity, and is nilpotent of order two, \emph{i.\ e.\ }
\begin{gather}
\delta^2 = 0.
\end{gather}
One can define \emph{another} BRST transformation $\bar{\delta}$, called anti-BRST, by
\begin{align}
\bar{\delta} A^{A}_{\mu} &= \partial_{\mu} \bar{C}^{A} + g f^{ABC} A^{B}_{\mu} \bar{C}^{C}, \nonumber \\
\bar{\delta} \bar{C}^{A} &= -\frac{1}{2} g f^{ABC} \bar{C}^{B}\bar{C}^{C}, \nonumber \\
\bar{\delta} \bar{C}^{A} &= i\bar{b}^{A}, \nonumber \\
\bar{\delta} \bar{b}^{A} &= 0,
\end{align}
where $\bar{b}^{A} = -b^{A} + igf^{ABC} C^{B} \bar{C}^{C}$.

Functions of the fields which are annihilated by  $\delta$ are called \emph{BRST closed}. A trivial subset of these are those functions which are themselves obtained by the action of  $\delta$ on some function; these are deemed \emph{BRST exact}. The BRST symmetry operator thus generates a co-homology, obtained as equivalence classes of BRST-closed functions modulo BRST-exact ones. According to standard arguments \cite{Henneaux:1992ig}, there exists a one-to-one correspondence between the BRST co-homology at ghost number zero and the set of classical physical observables (the gauge-invariant functions of the gauge fields).

It is to be stressed that no choice of gauge fixing is required to establish either the existence of the BRST transformation or of the correspondence between the BRST co-homology and gauge-invariant functions.
\section{BRST actions and gauge fixing}
\label{gfix}
A complete specification of a physical theory requires not only a set of physical observables (and their commutation relations for quantum mechanics), but also a prescription for the dynamics (via an action principle). In the context of the BRST formalism, it transpires that prescribing the action necessarily implies fixing the gauge.\footnote{One demands that the action unequivocally determines the trajectory in the extended phase space via the equations of motion, once boundary conditions are supplied. The gauge freedom, which allows the trajectories to be gauge transformed, must be abandoned. This is precisely gauge fixing.} In the original work \cite{Kondo:2001nq}, two classes of Lorentz-covariant gauge conditions are considered. The first class corresponds to the Lorentz gauge condition
\begin{gather} \label{lorg}
\partial_{\mu} A^{\mu} = 0.
\end{gather}
The gauge fixing is incomplete, or partial, in that there remains a subgroup of $SU(N)$ gauge transformations which preserve the gauge condition (\ref{lorg}).
A suitable BRST-closed Lagrangian corresponding to the gauge fixing in (\ref{lorg}) can be written as \cite{Curci:1976bt,Baulieu:1982sb,Baulieu:1985tg}
\begin{gather} \label{lors}
L = -\frac{1}{4} F_{\mu\nu}^{A} F^{A\mu\nu}
+i\delta \bar{\delta} (\frac{1}{2}A_{\mu}^{A}A^{A\mu} - \frac{i\alpha}{2} C^{A} \bar{C}^{A}).
\end{gather}
The second class of gauge conditions corresponds to the maximal Abelian gauge \cite{'tHooft:1981ht,Kronfeld:1987vd,Kronfeld:1987ri}. In this scenario, the gauge is fixed by choosing some operator $X$ which transforms in the adjoint representation, and performing a gauge transformation $V$ such that $X$ is diagonalized,
\begin{gather} \label{mag}
X \rightarrow \tilde{X} = V X V^{\dag} = \mathrm{diag} (\lambda_1, \dots, \lambda_N).
\end{gather}
Like the Lorentz gauge, this gauge fixing is also incomplete: $\tilde{X}$ maintains its diagonal form under the restricted set of gauge transformations belonging to the maximal torus $U(1)^{N-1}$ of $SU(N)$, \emph{viz.\ }
\begin{gather}
\{ d \epsilon SU(N) | d = \mathrm{diag} (e^{i\alpha_1}, \dots, e^{i\alpha_N}), \Sigma_i \alpha_i =0 \}.
\end{gather}
The gauge fields can be partitioned into $N-1$ diagonal components or `photons' $a^{i}_{\mu} = [A_{\mu}]_{ii}$ transforming as
\begin{gather}
a^{i}_{\mu} \rightarrow a^{i}_{\mu} + i\partial_{\mu} \alpha_i
\end{gather}
under the residual $U(1)^{N-1}$, and $N(N-1)$ off-diagonal `gluons' $c^{ij}_{\mu} = [A_{\mu}]_{ij}$ transforming as
\begin{gather} \label{cgt}
c^{ij}_{\mu} \rightarrow e^{i(\alpha_i - \alpha_j)} c^{ij}_{\mu}.
\end{gather}
The components of the gauge field $A^{A}_{\mu}$ in the basis $\{ T^A \}$ of generators of the Lie algebra can be written as linear combinations of the `photons' and `gluons'. This amounts to a change of basis. We denote the diagonal components by $\{ A^{i}| i=1,\dots, N-1\}$ and the off-diagonal components by $\{ A^{a} | a = N,\dots, N^2-1 \}$.

The gauge-fixed BRST-closed action can be derived from the Lagrangian \cite{Kondo:1998pc,Kondo:1998nw,Kondo:1998sr}
\begin{gather} \label{mags}
L = -\frac{1}{4} F_{\mu\nu}^{A} F^{A\mu\nu}
+i\delta \bar{\delta} (\frac{1}{2}A_{\mu}^{a}A^{a\mu} - \frac{i\alpha}{2} C^{a} \bar{C}^{a}),
\end{gather}
which is identical to (\ref{lors}), except that the gauge fixing part of the Lagrangian contains only the off-diagonal gauge fields.
\section{The mixed condensate}
\label{cond}
The mixed gluon-ghost condensate of dimension two proposed in \cite{Kondo:2001nq} is
\begin{gather} \label{O}
O = \frac{1}{\Omega} \int d^D x \; 
\mathrm{tr} \left[ \frac{1}{2} A_{\mu}  A^{\mu} - \alpha i C \bar{C} \right],
\end{gather}
where $\Omega$ is the volume of $D$-dimensional spacetime and the fields all take values in the Lie algebra of the gauge group $SU(N)$. The trace runs over the broken generators of the Lie algebra. Thus, in the Lorentz gauge, the trace runs over all generators, whereas in the maximal Abelian gauge, the trace runs only over the off-diagonal generators, \emph{i.\ e.\ }those in $SU(N)/U(1)^{N-1}$. 

The BRST invariance is shown by eliminating the $b$-fields via their equations of motion. In the Lorentz gauge for example, the $b$ fields satisfy 
\begin{gather} \label{b}
b^A = -\frac{1}{\alpha} \partial_{\mu} A^{A\mu} + i\frac{g}{2} f^{ABC} C^B \bar{C}^C.
\end{gather}
Once the auxiliary fields have been eliminated, the BRST transformation (\ref{BRSTeq}) reduces to the \emph{on-shell} form
\begin{align} \label{BRSTos}
\delta' A^{A}_{\mu} &= \partial_{\mu} C^{A} + g f^{ABC} A^{B}_{\mu} C^{C}, \nonumber \\
\delta' C^{A} &= -\frac{1}{2} g f^{ABC} C^{B}C^{C}, \nonumber \\
\delta' \bar{C}^{A} &=  -\frac{i}{\alpha} \partial_{\mu} A^{A\mu} - \frac{g}{2} f^{ABC} C^B \bar{C}^C.
\end{align}
The on-shell BRST variation of $O$ is thus zero,
\begin{gather} \label{inv}
\delta'  \int 
\left[ \frac{1}{2} A_{\mu}^{A}  A^{A \mu} - i \alpha  C^A \bar{C}^A \right]
= \int \partial_{\mu} (A^{A\mu} C^A) = 0.
\end{gather}
The BRST invariance of $O$ in the maximal Abelian gauge follows by replacing uppercase indices by lowercase ones in equations (\ref{b}-\ref{inv}), \emph{i.\ e.\ }restricting to the off-diagonal components.

Now we are confronted with the paradox. An operator has been found which is (on-shell) BRST-closed. It is clearly also BRST-inexact, and of ghost number zero. It is apparently, therefore, a \emph{bona fide} member of the BRST co-homology at ghost number zero and ought to correspond to a gauge-invariant operator. The correspondence states \cite{Henneaux:1992ig} that to obtain the gauge-invariant function from the BRST-closed operator, one simply discards the ghost fields. We thus have the apparent result that
\begin{gather} \label{gi}
O_0 = \frac{1}{\Omega} \int \mathrm{tr} A_{\mu} A^{\mu}
\end{gather}
is a gauge-invariant quantity! This is nonsensical.

To resolve this, let us go back to the BRST transformation written in its on-shell form (\ref{BRSTos}). To obtain this form, we have eliminated the $b$-fields via their equations of motion. But the Lagrangians (\ref{lors}) and (\ref{mags}), and thus the equations of motion, are \emph{dependent} on the gauge-fixing. Thus the on-shell BRST symmetry is somehow dependent on the gauge-fixing. But one \emph{still} has a BRST symmetry (\ref{BRSTos}),\footnote{It is straightforward to show that the on-shell action obtained by eliminating the $b$ fields is closed under the on-shell BRST transformation} which must correspond to \emph{some} gauge symmetry. What gauge symmetry? The answer must be that \emph{the on-shell BRST symmetry corresponds to the residual gauge symmetry which remains after the gauge has been partially fixed}.

Now the resolution of the paradox is clear. The operator $O$ is closed under the on-shell BRST transformation. It therefore has a corresponding gauge-invariant function. This gauge-invariant function need not be invariant under the full gauge group $SU(N)$, but only under the subgroup of gauge transformations which preserve the partial gauge-fixing. Let us now see this explicitly.

Consider performing an infinitesimal gauge transformation $U = 1 + O(\epsilon)$ which preserves the Lorentz gauge condition (\ref{lorg}). The variation of (\ref{gi}) is
\begin{align}
\int \mathrm{tr} A_{\mu} A^{\mu} 
&\rightarrow \int \mathrm{tr} (U A_{\mu} U^{\dag} + U \partial_{\mu} U^{\dag})^2 \nonumber \\
&= \int \mathrm{tr} (U  A_{\mu} A^{\mu}  U^{\dag} + 2 U A^{\mu} \partial_{\mu} U^{\dag} ) + O(\epsilon^2) \nonumber \\
&= \int \mathrm{tr} ( A_{\mu} A^{\mu} - 2 \partial_{\mu} A^{\mu} ) + O(\epsilon^2) \nonumber \\ 
&=  \int \mathrm{tr}  A_{\mu} A^{\mu},
\end{align}
where we have integrated partially and restricted to the subgroup of gauge transformations which preserve the gauge condition. The `gauge invariant' operator $O$ is thus invariant under the \emph{restricted} transformations which preserve the Lorentz gauge condition, but not the full gauge group.

Similarly, in the maximal Abelian gauge (\ref{mag}), the on-shell BRST closed operator of dimension two corresponds to the `gauge-invariant' quantity (\ref{gi})
\begin{gather}
O_0 = \int \mathrm{tr} A_{\mu} A^{\mu},
\end{gather}
where now the trace runs only over the off-diagonal generators of the Lie algebra. Performing a change of basis, one has that
\begin{gather}
O_0 = \int c_{\mu}^{ij}c^{ji\mu}.
\end{gather}
Under a $U(1)^{N-1}$ gauge transformation (\ref{cgt}), it is straightforward to show that $O_0$ is invariant, as required.

\section{Discussion}
The status of the `BRST-invariant ' operator of mass dimension two is now clear. It is not strictly BRST-closed; rather it is closed under an on-shell BRST symmetry which corresponds to the residual gauge freedom which remains once one has performed a partial gauge fixing. This raises an important point regarding the BRST construction in the general case. It is quite common to eliminate fields via their equations of motion, for example, when passing from the Hamiltonian to the Lagrangian formalism. The BRST symmetry which results will only correspond to the full gauge symmetry if no gauge-fixing is invoked in eliminating the fields. In practice this means that one is only free to eliminate those fields which are not coupled to the so-called `gauge-fixed' part of the action.

Given that the BRST symmetry of the mixed condensate corresponds only to a \emph{subset} of gauge symmetries of the theory, it is clear that the condensate is no more or less useful than other gauge-variant quantities. In a fixed gauge, the condensate (\ref{O}) is a correct BRST-invariant extension of the gauge-invariant operator (\ref{gi}).  Provided one sticks to the chosen gauge, the condensate is BRST-invariant, and calculations are perfectly valid. However, the problem of how to extend the condensate off the chosen gauge slice in a gauge- (or rather BRST-) invariant fashion still remains. 

Clearly what would be useful is a dimension-two operator which is closed under the full BRST transformation (\ref{BRSTeq}). But this is no easier to write down than a fully gauge-invariant function of dimension two. The only possibilities for progress appear to be the consideration of non-local quantities \cite{Gubarev:2000eu}, or the possibility that physics may be hidden in the co-homology at non-zero ghost number, which is not understood.
\section{Acknowledgement}
I would like to thank I. I. Kogan for discussions.

\end{document}